\documentstyle[preprint]{aastex}

\begin{document}

\title{The Ratio of Blue to Red Supergiants in Sextans A from HST
Imaging
\footnote{based on observations with the NASA/ESA Hubble Space Telescope
obtained at the Space Telescope Science Institute, which is operated by
the Association of Universities for Research in Astronomy, under NASA
contract NAS 5-26555.}}

\author{Robbie C. Dohm-Palmer}
\affil{University of Michigan, Department of Astronomy, Ann Arbor, MI
48109}
\email{rdpalmer@astro.lsa.umich.edu}

\author{Evan D. Skillman}
\affil{University of Minnesota, Department of Astronomy, Minneapolis, MN
55455}
\email{skillman@astro.umn.edu}

\begin{abstract}

We have examined the ratio of blue to red (B/R) supergiants in the
dwarf irregular galaxy Sextans~A. The supergiants were identified in
previously published stellar photometry measured from Hubble Space
Telescope imaging. The high resolution imaging and low dust
environment provided high photometric accuracy such that the main
sequence and blue He-burning supergiants are clearly separated. This
allows us to isolate the He-burning phase at both the red and blue
ends of the so called ``blue-loops''. The B/R supergiant ratio
provides an observational constraint on the relative lifetimes of
these two phases which is a sensitive test for convection, mass loss,
and rotation parameters. These parameters have direct implications for
the period-luminosity relationship for Cepheid variable
stars. Previous studies have used a single number to represent this
ratio. However, since the B/R ratio is a fairly strong function of
mass for a single age stellar population, both changes in recent 
star formation rate and choice of luminosity cut-off can dramatically
affect the result.  We have analyzed the ratio as a
function of age, or equivalently, mass.  This method eliminates the
confusion of unknown star formation histories so that B/R can be a
more reliable diagnostic tool.  We compare the result with a model
based on stellar evolution tracks of an appropriate metallicity. The
functional form of the observed ratio matches the model extremely
well. However, the observed B/R ratio is lower than the model by a
factor of two.  This result suggests that stellar rotation is an
important effect in the evolution of these stars.

\end{abstract}


\section{Introduction}

The ratio of blue supergiants to red supergiants (B/R) is an important
diagnostic for stellar evolution models. In particular, this ratio is
a sensitive test for convection, mass loss, and rotation parameters
(see \citet{lan95} and \citet{mae01} for summaries; see also
\citet{sal99}).  As \citet{lan95} discuss, some models 
reproduce the observed ratios for low metallicity, and others do
so for high metallicity. However, there is not yet a self-consistent
model that can reproduce (B/R) at both high and low metallicity. The
effects of mass loss and rotation will also affect the period of
variable stars \citep{mae01}.  Thus, any application of, for example,
the Cepheid period-luminosity relationship must account for these
parameters, and their relationship to metallicity.

Observational studies have been aimed at measuring absolute values of
the B/R ratio and also measuring this ratio as a function of
metallicity.  Previous studies have measured the B/R ratio based on
all stars brighter than a chosen absolute magnitude limit (see
references in \citet{lan95}). Typically the calculation was restricted
to very massive stars, ($> 15 M_\sun$). This provided a single ratio
for each stellar population. For many of these studies this was the
only reasonable method because of the limited photometric accuracy.
However, there are several theoretical difficulties with this
approach.

First, it has been known for some time that the B/R ratio is a fairly
strong function of luminosity, and thus mass (e.g., \citet{sto69};
\citet{lan95}).  This makes comparison between galaxies difficult.
For the comparison to be valid, the mass distribution of the
supergiants would have to be identical above the magnitude limit.
In other words, the star formation histories (SFHs) would have to be
identical.

Second, it is obvious that the magnitude limits would need to be
identical, but due to different distances of objects, this is not
always practical.  Additionally, when comparing galaxy populations to
cluster populations, clusters almost never have a statistically
significant population of very massive stars, so that the B/R ratios
reported for clusters almost always use fainter limiting magnitudes.

Third, in ground-based observations of extragalactic objects, the blue
supergiants are usually photometrically confused with the main
sequence (MS). Thus, the number of blue supergiants includes those MS
stars brighter than the chosen limit.  Clearly these MS stars do not
have red supergiant counterparts, producing a bias to higher B/R
ratios which are not directly comparable to ratios derived from
stellar evolution models.  If the contamination by MS stars is
accounted for in the models, again, the SFH has a strong effect on the
resulting B/R ratio.

Clusters are not immune to these difficulties, even though they are
isochronic.  Since clusters have different ages, each cluster will
have supergiant stars with masses corresponding to its particular age.
Hence, valid comparison can only be made between clusters with the
same age.

To properly use observations to constrain models, one must either
incorporate the SFH into the model or eliminate this dependence from
the observations. The most reliable way to eliminate the SFH from
observations is to compare blue and red stars of the same age. In
galaxies with mixed age populations the stars can be divided into age
groups. Thus, one can calculate a function, or histogram, of the B/R
ratio as a function of age, rather than a single number for the
galaxy. In the following section we briefly describe the data and our
method for calculating the B/R ratio. We compare the result with a
model calculation. The final section discusses the results and the
implications for stellar evolution models.  Here we will use Hubble
Space Telescope observations of Sextans A to demonstrate this method.

\section{The B/R Supergiant Ratio in Sextans A}

HST observations of the dwarf irregular galaxy Sextans~A provided, for
the first time, a clear photometric separation of MS stars from
post-MS supergiants in their core He-burning (HeB) phase of evolution
\citep{doh97a}.  Using the isolated blue HeB stars, a detailed SFH was
determined over the past 600 Myr \citep{doh97b}.  These data were
later enhanced by additional HST observations \citep{doh01}. Combining
the two data sets provides nearly complete coverage of the optically
visible portion of the galaxy. It also provides enough supergiants to
perform a statistically feasible analysis of the B/R supergiant ratio.
The combined data are shown for the upper part of the color magnitude
diagram in Fig.\ \ref{figcmd}.

   From these data sets we have isolated the blue and red supergiants in
the color-magnitude diagrams (CMDs). For the magnitude range of
interest, the MS and BHeB populations show little or no change in
color-index. Furthermore, these two populations are well separated
compared to the photometric errors. To demonstrate this, we have
calculated a histogram in color-index (Fig.\ \ref{fighist}) for the MS
and BHeB stars with $-5.6 < M_V < -3.2$. The histogram shows an
indisputable separation of these two populations. To determine the
color-index selection limit for the BHeB stars, we simultaneously fit
two Gaussian curves to the histogram. The result is shown in Fig.\
\ref{fighist}. The equality point between the two fit curves is
(V-I)$_0 = -0.13$. We selected the BHeB stars as those stars redward
of this limit.

   From the simultaneous fit in Fig.\ \ref{fighist} we can also estimate
the degree of contamination of MS stars into the selected BHeB
population. We integrated the curves of those fits and found that 5\%
of the BHeB stars are blueward of the selection limit. We also found
that 6\% of the of the BHeB counts are actually due to MS stars. Thus
the net loss of stars from the blue count is approximately 1\%. This
is much smaller than the Poisson errors. 

The red stars were selected interactively. Brighter than the RGB, the
red supergiants are well isolated from other populations, so the exact
selection region in the CMD is not important. The faint end of the red
supergiants blends with both the AGB and the RGB. This sets a
practical age limit for this calculation of $\sim200$ Myr.  Even
though the observations in \citet{doh01} are much deeper than those in
\citet{doh97b}, the errors in the photometry and the completeness
corrections for the supergiants of interest here are negligible.

We calculate the B/R ratio in age bins, rather than as a single number
for the galaxy. By comparing blue and red stars of the same age we
eliminate any dependence on the star formation rate (SFR). The SFR can
in general, and does in Sextans~A \citep{doh01}, change over the time
periods for which supergiants exist. It is also crucial to separate
the blue supergiants from the MS.  MS stars and blue supergiants of
the same magnitude have very different masses and ages. Thus, if these
populations are mixed, the calculated ratio is extremely sensitive to
both the star forming history as well as the initial mass function
(IMF).

By separating the stars into age bins, the risk of contamination
of the BHeB counts by MS stars 
is somewhat enhanced.  For example, if, at a given luminosity, the
number of MS stars greatly exceeds the number of BHeB stars, then
the $\sim$5\% of MS stars that will contaminate the BHeB stars 
could have a disproportionate effect on the BHeB star count in 
that bin.  We have investigated this by plotting histograms similar
to Figure 2 for each age bin, and find that the number of MS stars
and BHeB stars is roughly comparable in each age bin.  In many
bins there is a clear gap between the two populations.  Thus the
contamination of MS stars into the BHeB star counts should be 
comparable to the number of BHeB stars lost to MS star counts
in each bin.  Overall, the effect of confusion between the MS and
BHeB star counts is conservatively estimated to be less than 10\% 
in each bin.  That is, this will be a negligible effect on the
outcome.

Table \ref{tabcum} shows the B/R ratio if we were to adopt a faint
magnitude limit and calculate a single number for the galaxy. The
first column is the adopted magnitude limit, and the second is the
resulting B/R ratio. We include several magnitude criteria to
demonstrate the dependence of this ratio on the adopted limit. The
ratio generally decreases with fainter limits.  We see here nearly a
factor of two difference between the B/R ratio obtained for the most
luminous stars (comparable to galaxy samples) and that obtained for
stars with M$_V$ $\le$ $-$3 (comparable to cluster samples).

To assign ages to the selected stars we have used the stellar
evolution models of \citet{sch92}, combined with the stellar
atmosphere models of \citet{lej01}.  We will refer to these as the
Geneva models. We used the Z=0.001 models to match the nebular
abundance of Sextans~A \citep{ski98}. The models of this metallicity
have been shown to be a good match to the stellar population
\citep{doh01}, with the blue and red He-burning sequences in the
correct predicted locations (see Fig.\ \ref{figcmd}). For both the
blue and red supergiants, each star can be assigned a nearly
unambiguous age based on its position in the CMD. This is possible for
two reasons. First, the lifetimes in these phases are very short
compared to the age in which they are entered. Second, the luminosity
of these phases is nearly a monotonic function of mass, and hence age.

To be complete, we have also applied a correction for the IMF. We used
a power law function with a Salpeter slope of -1.35 \citep{sal55}. For
a given age, the mass of a blue supergiant will be slightly different
from the mass of a red supergiant. However, the mass difference is so
small that the resulting correction is negligible. We experimented
with a wide range of IMF slopes, but found the B/R ratio to be
insensitive to this parameter.

The result of the B/R ratio calculation is show in Fig.\
\ref{figratio}. The calculated values are given in Table
\ref{tabbr}. Note that the ages listed are the lower bounds of the
histogram bins. The observed ratio is indicated with a histogram. The
ratio starts near 2.5 and decreases with age. We have limited the
calculation to ages older than 20 Myr. The evolution of younger, more
massive, stars is highly uncertain (see \citet{doh97b} for a
discussion).

To determine the predicted ratio we have created a simulated CMD using
the Z=0.001 Geneva stellar evolution models. The simulated CMD was
created using Monte Carlo techniques outlined by several authors
(e.g., \citet{tos91}; \citet{tol96}). We did not incorporate
photometric errors into this model. This would only serve to smooth
the B/R ratio function in age, and we preferred a pristine model
prediction. We also did not apply an incompleteness correction. The
observed data is complete for the relevant populations.

There were 200,000 stars in the model, distributed with a power law
IMF with a Salpeter slope (-1.35; \citet{sal55}). The model used a
constant star formation rate over the past 200 Myr. This resulted in
5502 blue supergiants and 3843 red supergiants. These stars were then
divided into age bins in exactly the same manner as the observed data,
including the IMF correction. The model prediction is included in
Fig.\ \ref{figratio} as a solid thick line.

The predicted ratio is approximately a factor of 2 larger than the
observed ratio for all ages. However, the functional forms of the
model and observations are remarkably similar. To emphasize this, we
have plotted the model prediction reduced by a factor of two as a
dashed line.

\section{Errors in the Stellar Evolution Models}

In the previous section we claimed to have eliminated the uncertainty
of the star formation history from the calculation of the B/R
ratio. Strictly speaking, this is only true if the adopted
luminosity-age relationship is accurate for both the red and blue
stars. We wish to emphasize three points concerning this issue.

While the stellar evolution models available today are not expected
to be perfect, we have found that, in many respects, they are quite 
good in reproducing the observations.  For example, \citep{doh97b}
showed that the stellar evolution models (from both the Geneva and
Padua groups) of the appropriate metallicity reproduced the positions 
of the blue and red supergiants in the V, V-I color magnitude diagrams
over their entire extent (also shown here in Figure 1).  Given the
uncertainties in these models (and the lack of very low metallicity,
young stellar clusters with which to calibrate the models), we find 
this agreement both surprising and reassuring.

Second, we have found that the model predictions for the relative
ages of the BHeB and MS stars agree remarkably well with the 
photometric observations in Sextans~A \citep{doh01}.  That is, we have
compared the star formation histories calculated from MS stars and
BHeB stars, and these two calculations show excellent agreement both
qualitatively and quantitatively.  The excellent agreement in the 
functional form of the star formation histories, which is only
dependent on the adopted age-luminosity, and mass-luminosity
relationships, gives us great confidence that, at least for the blue
stars, these relationships are reliable.

Third, and perhaps more important, is that the present calculations are
designed to constrain stellar evolution models.  Again, we recognize
that evolution models are not perfect, and that there is still much work 
to be done. However, that should not prevent us from using the available
models to calculate quantities such as the star formation
history.  Such calculations, even given the possible inaccuracies due
to less than perfect stellar evolution models, can still teach
us much about the objects we study.  We must simply accept the
uncertainty as being the best one can do at present, just as we must
accept the uncertainty in other calculations, such as distance
estimates.

In the specific case of this paper, we have used stellar evolution
models to self-consistently calculate the B/R ratio as a function of
age. We find that the observed ratio does not match the predicted
ratio. This implies that the model needs to be adjusted in some
manner. It is possible that either, or both, the luminosity-age
relationship and the relative lifetimes of the two phases need
adjustment. 

\section{Discussion}

The stars of Sextans~A show that, in the specific case of the Geneva
models, the predicted B/R ratio is a factor of 2 too large for this
metallicity.  However, these models do not incorporate stellar
rotation. It has been shown by \citet{mae01} that rotation increases
the core mass during the He-burning phase. This inflates the
atmosphere, increases mass loss, and pushes stars redward. Similar
results have been found by \citet{sal99} by increasing mass loss in
the red supergiant phase. Thus, a smaller B/R ratio is predicted by
incorporating rotation \citep{mae01}. This is exactly the effect
needed to account for our observations. It should be noted that the
stars discussed by \citet{mae01} have masses larger than the masses we
consider here, however, the same general trends should apply.

Previous measurements of the B/R ratio have produced a single number
for each population studied.  As discussed in the introduction, such
measurements can be deceiving, and comparisons with models must be
made with care. For field stars, the star forming history influences
this ratio making comparisons with other objects, and models,
difficult. Even for isochronic populations, such as clusters, care
must be taken to compare the result with a model of the appropriate
age.

\citet{lan95} discuss the dependence of the B/R ratio on
metallicity. The observational data they discuss suggest that this
ratio increases with increasing metal abundance. This trend may in
fact be true, however, the observational data points have not been
adequately decoupled from age and star formation history. We have not
resolved the problem of models not matching the B/R ratio at all
metallicities.  However, we have presented a method for removing
ambiguities from the observed B/R ratio, making comparisons with
models more meaningful.  New calculations of stellar evolution
including rotation at a metallicity appropriate to Sextans A would
certainly be of interest.

We have demonstrated a technique for deriving a reliable constraint to
evolution models of massive stars.  This technique relies on clearly
separating the blue supergiants from the MS.  Unfortunately this may
limit the number of objects available for analysis. High resolution
imaging, such as with the HST, is required for objects with distance
of order 1 Mpc or further.  Additionally, in higher metallicity
environments, dust can significantly broaden these features in the
color magnitude diagram.  In which case, multi-color photometry will
be required to correct for line-of-sight reddening effects.

\acknowledgments

Support for this work was provided by NASA through grant GO-7496 from
the Space Telescope Science Institute, which is operated by AURA,
Inc., under NASA contract NAS 3-26555.  Partial support from NASA
LTSARP grant no.\ NAG5-9221 and the University of Minnesota is
gratefully acknowledged.  This research has made use of NASA's
Astrophysics Data System Abstract Service. 

We thank the referee for insightful comments that improved this
manuscript.

\begin{deluxetable}{rr}
\tablecaption{B/R Ratio Based on a Magnitude Limit}
\tablewidth{0pt}
\tablehead{\colhead{$M_V$} & \colhead{B/R ratio}}
\startdata
$-$6.0 & 3.25 \\
$-$5.5 & 2.70 \\
$-$5.0 & 2.78 \\
$-$4.5 & 2.61 \\
$-$4.0 & 1.77 \\
$-$3.5 & 1.76 \\
$-$3.0 & 1.85 \\
\enddata
\label{tabcum}
\end{deluxetable}

\begin{deluxetable}{rr}
\tablecaption{The B/R ratio as a Function of Age}
\tablewidth{0pt}
\tablehead{\colhead{Age (Myr)} & \colhead{B/R ratio}}
\startdata
20  & 2.55 \\
35  & 1.54 \\
50  & 0.70 \\
65  & 1.28 \\
80  & 0.90 \\
95  & 0.73 \\
110 & 0.66 \\
125 & 0.43 \\
\enddata
\label{tabbr}
\end{deluxetable}

\begin{figure}
\caption{The color-magnitude diagram for Sextans~A. The stellar
evolution tracks for the adopted Z=0.001 Geneva models (see text) are
also plotted. These are labeled with the starting mass in Solar
units. Notice that the ends of the blue loops in the models align with
the observed blue and red supergiant sequences. We have indicated the
age bins for the blue and red supergiant sequences on the left and
right respectively. Notice that the age-luminosity relationship for
the blue and red stars is different. Thus, limits based on magnitude
alone would result in mixing different age populations.}
\label{figcmd}
\end{figure}

\begin{figure}
\caption{A color-index histogram of Sextans~A stars with absolute V
magnitude between -5.6 and -3.2. This is the magnitude range used in
the B/R ratio calculation. The MS and BHeB populations are clearly
well separated. We have simultaneously fit two Gaussian curves to the
histogram. The separate fits are shown as dashed lines, and the sum
of the two is shown as a solid line. We have also indicated the
equality point with a vertical line at (V-I)$_0 = -0.13$.}
\label{fighist}
\end{figure}

\begin{figure}
\caption{The B/R ratio for the dI galaxy Sextans ~A. The histogram is
the observed ratio as a function of age. Stars were divided into age
bins using the stellar evolution models of \citet{sch92}. The error
bars reflect Poisson errors. The thick line is the ratio determined
from a Monte Carlo CMD model using the stellar evolution models of
\citet{sch92}, with atmospheric models applied by \citet{lej01}. The
error bars also reflect Poisson errors. The dashed line is the model
reduced by a factor of 2. Notice how well the functional form of the
observations matches that of the model. However, the model values are
twice as large as the observations.}
\label{figratio}
\end{figure}


\begin{thebibliography}{}

\bibitem[Dohm-Palmer et al.\ (1997a)]{doh97a} Dohm-Palmer R.C., Skillman
E.D., Saha A., Tolstoy E., Mateo M., Gallagher J., Hoessel J., Chiosi
C. \& Dufour R.J. 1997a, \aj, 114, 2514

\bibitem[Dohm-Palmer et al.\ (1997b)]{doh97b} Dohm-Palmer R.C.,
Skillman E.D., Saha A., Tolstoy E., Mateo M., Gallagher J., Hoessel
J., Chiosi C. \& Dufour R.J. 1997b, \aj, 114, 2527

\bibitem[Dohm-Palmer et al.\ (2001)]{doh01} Dohm-Palmer R.C., Skillman
E.D., Mateo M., Saha A., Dolphin A., Tolstoy E., Gallagher J.S., \&
Cole A.A. 2001, \aj, in press

\bibitem[Langer \& Maeder (1995)]{lan95} Langer N. \& Maeder A. 1995,
\aap, 295, 685

\bibitem[LeJeune \& Schaerer (2001)]{lej01} Lejeuene T. \& Schaerer
D. 2001, \aap, 366, 538

\bibitem[Maeder \& Meynet (2001)]{mae01} Maeder A. \& Meynet G. 2001,
\aap, 373, 555

\bibitem[Salasnich, Bressan, \& Chiosi (1999)]{sal99} Salasnich B.,
Bressan A., \& Chiosi C. 1999, \aap, 342, 131

\bibitem[Salpeter (1955)]{sal55} Salpeter E.E. 1955, \apj, 121, 161

\bibitem[Schaller et al.\ (1992)]{sch92} Schaller G., Schaerer D.,
Meynet G., \& Maeder A. 1992, A\&AS, 96, 269

\bibitem[Skillman, Kennicutt, \& Hodge (1989)]{ski98} Skillman E.D.,
Kennicutt R.C., \& Hodge P.W. 1989, \apj, 347, 875

\bibitem[Stothers (1969)]{sto69} Stothers R. 1969, \apj, 155, 935

\bibitem[Tolstoy \& Saha (1996)]{tol96} Tolstoy E. \& Saha A. 1996,
\apj, 462, 684

\bibitem[Tosi et al.\ (1991)]{tos91} Tosi M., Greggio L., Marconi G.,
\& Focardi P. 1991, \aj, 102, 951

\end{thebibliography}
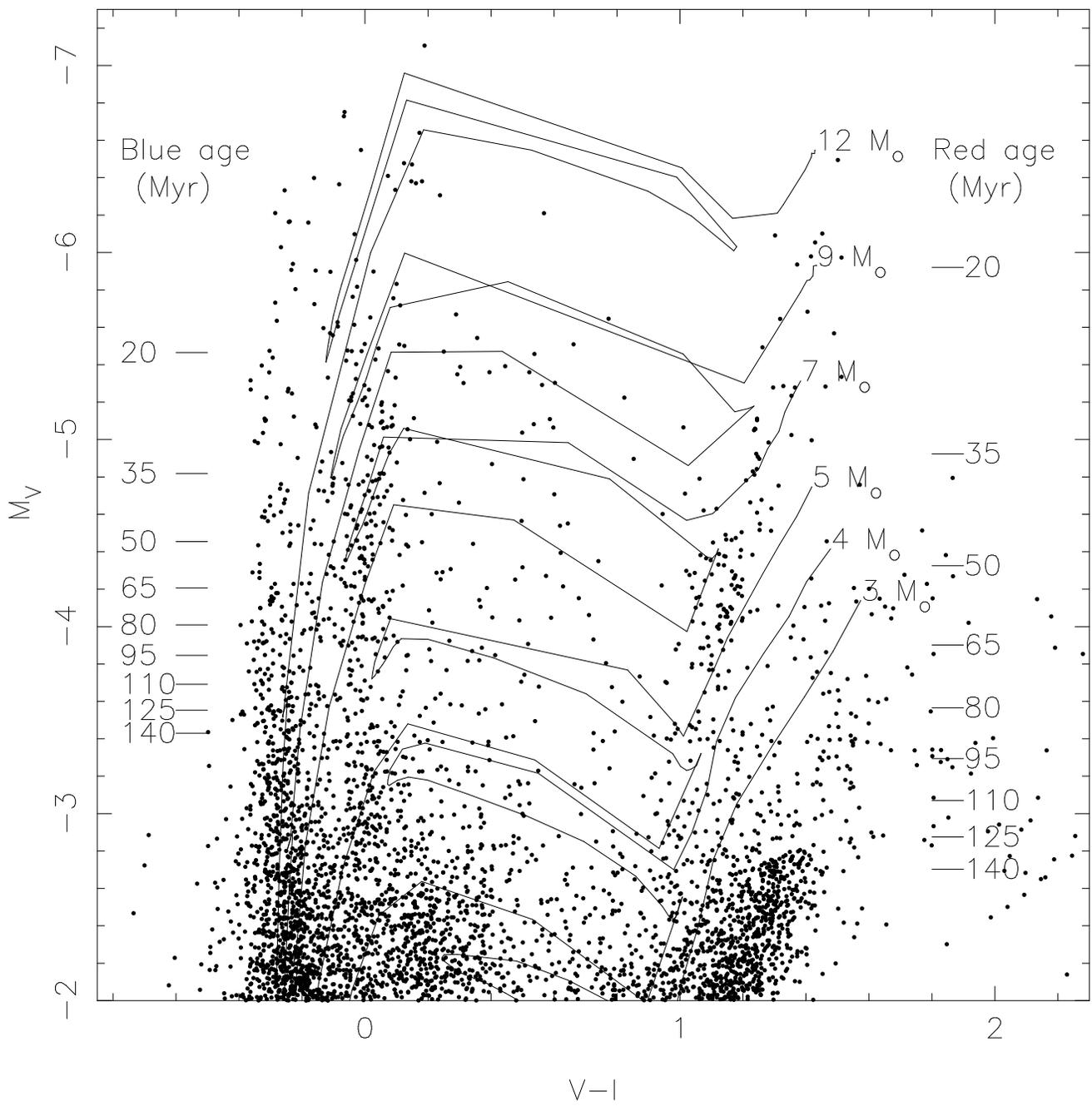
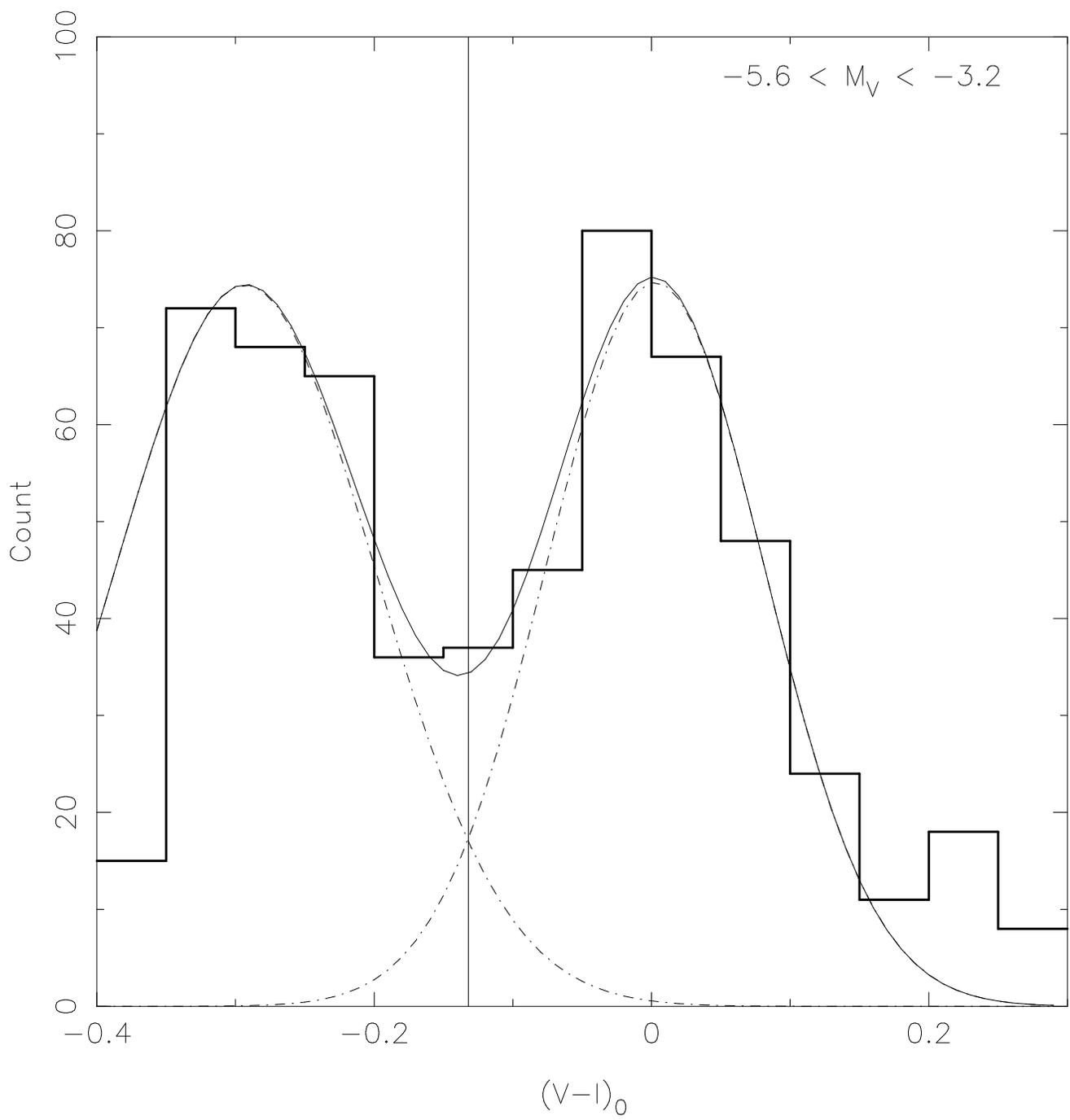
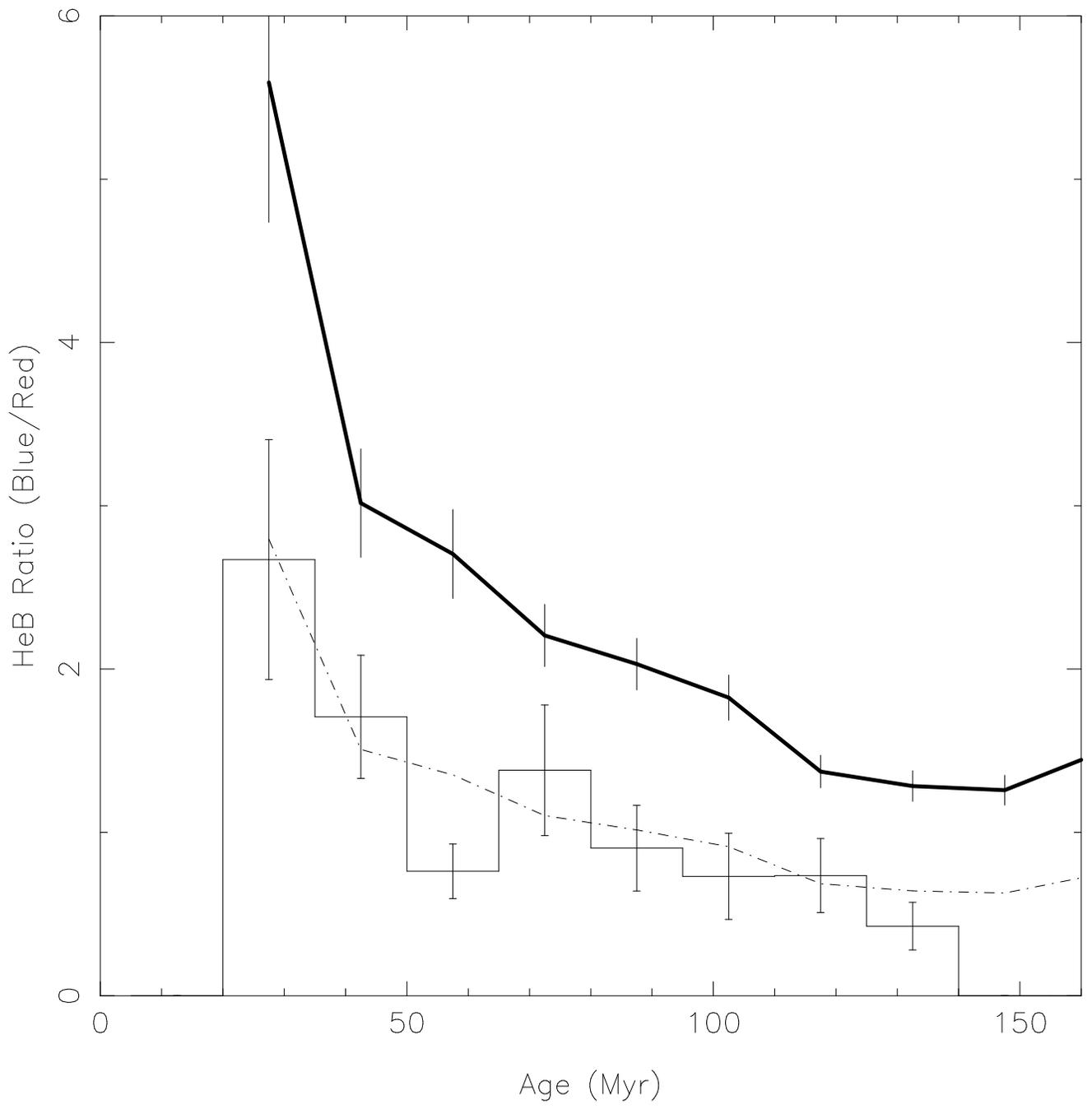
\end{document}